# Application of entropy analysis in the prediction of flow distribution in parallel channels

Toochukwu Aka, Shankar Narayan*

Department of Mechanical, Aerospace, and Nuclear Engineering, Rensselaer Polytechnic Institute, 110 8th Street, Troy, NY 12180

*Corresponding Author: narays5@rpi.edu

**ABSTRACT**

Multiphase flow in parallel channels is often an efficient approach to manage heat and energy distribution in engineering systems. However, two-phase flow with heating in parallel channels is prone to maldistribution, resulting in sub-optimal performance and in some cases, permanent damage. This challenge requires accurate flow modeling in parallel channels to mitigate or design against the adverse effect of two-phase flow maldistribution. The nonlinear nature of multiphase flow results in a multiplicity of predicted solutions for the same condition, thereby creating significant challenges in modeling flow distribution. Therefore, this study focuses on solving this challenge by applying entropy generation analysis and the conservation of mass, momentum balance, and energy balance to predict two-phase flow distribution in a two-parallel-channel assembly with a numerical model. Both model predictions and experimental data show that equally distributed flow becomes severely maldistributed with a decrease in flow rate, resulting in significant change (>30%) in the entropy generation rate. We show that the entropy analysis can be applied in distinguishing between stable and unstable flow distribution, like the linear stability analysis used in previous studies. We also surpass the limit of applying linear stability analysis by using entropy analysis to identify the most feasible end state in a maldistribution process.

# 1 INTRODUCTION

Flow distribution is critical to multi-channel engineering systems, ranging from heat exchangers and cooling systems to microfluidics and fuel cells. In multi-channel heat exchangers, flow distribution influences the contribution of each channel to heat transfer and the overall heat transfer efficiency[1][2] In microfluidics, precise flow distribution is vital for sample manipulation, precise dosing, and efficient reactions[3]. In fuel cells, the distribution of the reactants among parallel flow channels affects electrochemical efficiency and cell lifetime[4]. However, accurately predicting and understanding two-phase flow distribution in parallel channels presents significant challenges.

Several prior studies have been dedicated to analyzing and controlling flow distribution in parallel channels. In our previous computational study[5], we showed that the thermophysical properties of the channel walls can significantly influence flow maldistribution in two parallel channels. Zhang et al.[6] presented a linear stability analysis to distinguish between stable and unstable flow distributions in a multi-channel evaporator. In Zhang's study, a feedback control strategy was developed to maintain near-equal fluid distribution in a three-parallel channel assembly. Taitel et al. [7] introduced finite disturbances to demonstrate the stability of transient flow distribution solutions. Minzer et al.[8] also performed a linear stability analysis on static flow distribution solutions and showed that flow distribution in a parallel channel assembly depends on the history of the inlet flow rate.

Linear stability analysis was commonly applied in previous studies to determine the stability of flow distributions[6], [8]. However, it provides no physical insight into why a stable flow distribution is preferred over other "mathematically feasible" distributions. Also, linear stability analysis cannot be applied when multiple stable distributions correspond to a given operating condition. We address these limitations by adopting a thermodynamic approach to analyzing flow distribution in a parallel channel system.

Entropy analysis provides valuable insight into the direction and inefficiencies of physical processes in a system. Based on the second law of thermodynamics, entropy generation quantifies the rate at which entropy is produced during a physical process. Hence, previous studies have applied entropy analysis in design optimization[9]–[12], flow regime identification[13] and alternative approaches to known phenomena[14]. In this study, we conduct an entropy analysis for a two-parallel-channel assembly to show the relationship between flow distribution and the entropy production rate. We use entropy generation to explain the preference for stable over unstable flow distributions. Based on the characteristics of entropy generation in individual channels and common headers of the assembly, we show that this approach can be used in predicting the most feasible stable final states in processes prone to maldistribution.

## 2 ANALYSIS

### 2.1 Physical system

This study focuses on a two-parallel-channel assembly with a common inlet and exit, as shown in Figure 1. Each channel branch consists of a valve and a long steel tube (0.3048 m) with steady and uniform heating. Each valve has a flow coefficient, $K_v$ of $10^{-8}$, with an orifice opening, $A_v$ ranging from 0 to 100%. Subcooled water (working fluid) enters through the common inlet at $T_i =$19 °C and exits the parallel channel assembly as either liquid, liquid-vapor mixture, or superheated vapor at $P_e = 20$ kPa, while heat is transferred from the heaters to the working fluid.

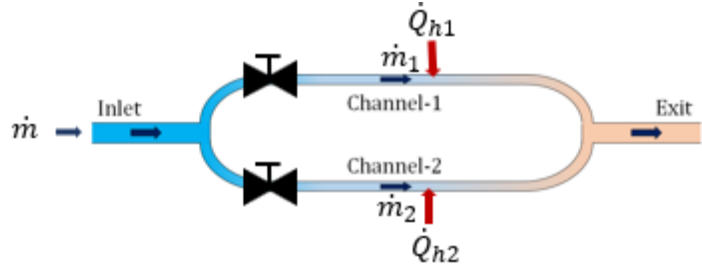

Figure 1. Two thermally isolated parallel channels sharing a common inlet and exit.

## 2.2 Governing Equations

The evolution of multiphase flow within a heated channel (Figure 1) can be described by the lumped form of the unsteady mass (Eq. (1)), momentum (Eq. (2)), and energy balance (Eq. (3) and Eq. (4)) equations.

$$A_{cs}\left(L\frac{d\rho}{dt}\right)_{ph} = (\dot{m}_i - \dot{m}_e)_{ph} \tag{1}$$

$$\frac{L}{A_{cs}}\frac{d\dot{m}}{dt} = P_i - P_e - \Delta P \tag{2}$$

$$A_{cs}\frac{d(\rho h - P)_{ph}}{dt} = pH_{ph}(T_w - T)_{ph} + \left(\frac{\dot{m}_i h_i - \dot{m}_e h_e}{l}\right)_{ph} \tag{3}$$

$$\rho_w c_{p,w}\left(V_w \frac{dT_w}{dt}\right)_{ph} = \left(\dot{Q}_h - Hpl(T_w - T) - \dot{Q}_{loss}\right)_{ph} \tag{4}$$

$\dot{m}_{ph}$ is the average mass flow rate across each fluid phase region (subcooled liquid, liquid-vapor mixture, superheated vapor) in the channel with subscript $i$ denoting inlet and $e$ denoting exit of the region. $\rho_{ph}$, $h_{ph}$, $P_{ph}$, $T_{ph}$, $H_{ph}$, and $l_{ph}$ are the average fluid density, enthalpy, pressure, temperature, and convective heat transfer coefficient for each phase, respectively. $T_{w,ph}$ and $V_{w,ph}$ describe the average temperature and volume of the wall corresponding to each phase in the channel. Flow properties related to channel geometry, specifically $L$, $p$ and $A_{cs}$ are the channel length, wetted perimeter, and flow cross-sectional area, respectively. Thermophysical properties of the channel wall $\rho_w$ and $c_{p,w}$ are the density and specific heat capacity, respectively. The pressure drop, $P_d$ across a channel branch consists of the valve ($\Delta P_v$), flow acceleration ($\Delta P_a$) and frictional ($\Delta P_f$) components.

$$\Delta P = \Delta P_v + \Delta P_a + \Delta P_{f,liq} + \Delta P_{f,tp} \tag{5}$$

$$\Delta P_v = \frac{1}{\rho_i}\left(\frac{\dot{m}}{K_v A_v}\right)^2 \tag{6}$$

$$\Delta P_a = \dot{m}^2 \left(\frac{1}{\rho_e} - \frac{1}{\rho_i}\right) \tag{7}$$

where $A_v$ is the valve opening, $\rho_i$ and $\rho_e$ are the average fluid density at the inlet and exit, respectively. The pressure drop for the liquid phase region $\Delta P_{f,liq}$ is given by the Darcy-Weisbach equation [5] with friction factor obtained from correlation in previois study[15], while the pressure drop in the two-phase region $\Delta P_{f,tp}$ is computed using the Lockhart and Martinelli correlation [8]. The average heat transfer coefficient for the liquid phase region is given by

$$H_{liq} = \frac{k_{liq}\, Nu_{liq}}{D} \tag{8}$$

where $k_{liq}$, $Nu_{liq}$ and $D$ are the average fluid thermal conductivity, Nusselt number, and channel internal diameter, respectively. $Nu_{liq}$ is given based on the assumption of a uniform heat flux [16]. The heat transfer coefficient in the two-phase flow regions and the critical heat flux (CHF) are computed using correlations from prior publications[17], [18]. Apart from the added simplification of a uniform channel heat flux, we adopt $P_e$ as the reference pressure for computing saturated fluid properties. Accordingly, the phase length of the liquid region in a channel, $l_{liq}$ is given by.

$$l_{liq} = \left(\frac{h_{liq,sat}(P_e) - h_i}{h_e - h_i}\right) L \tag{9}$$

The heat loss to the ambient $\dot{Q}_{loss}$ is obtained from experimental data as a function of $T_w - T_\infty$, and the outer surface area $(A)$, as shown in Figure 2. For model validation, heat loss to the surroundings takes place at an ambient temperature of $T_\infty = 24\ °C$.

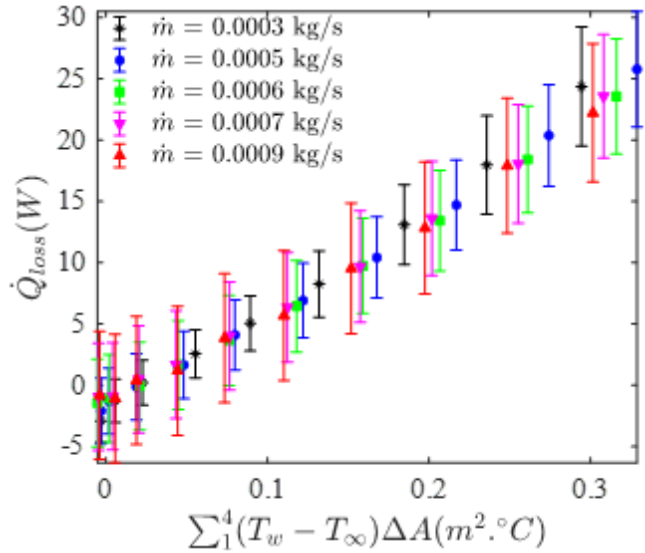

Figure 2. Heat loss characterization.

### 2.2.1 Static Model

The unsteady terms in Eqs. (1) to (4) can be eliminated at steady operating conditions, resulting in the steady forms of mass conservation, momentum balance, and energy conservation equations.

$$\dot{m}_i = \dot{m}_e = \dot{m} \tag{10}$$

$$P_i - P_e = \Delta P \tag{11}$$

$$p\left[l_{liq}H_{liq}(T_w - T)_{liq} + l_{tp}H_{tp}(T_w - T)_{tp}\right] = \dot{m}(h_e - h_i) \tag{12}$$

$$\dot{m}(h_e - h_i) = \dot{Q}_h - \dot{Q}_{loss} \tag{13}$$

And the steady rate of heat transfer $\dot{Q}_i$ into the fluid may be expressed as

$$\dot{Q}_i = \dot{m}(h_e - h_i) \tag{14}$$

The average wall temperature $T_w$ and fluid temperature $T$ is given by the following equations.

$$T_w = \left(T_{liq} + \frac{\dot{Q}_i}{H_{liq}pL}\right)\frac{l_{liq}}{L} + \left(T_{tp} + \frac{\dot{Q}_i}{H_{tp}pL}\right)\frac{l_{tp}}{L} \tag{15}$$

$$T = \frac{T_{liq}l_{liq} + T_{tp}l_{tp}}{L} \tag{16}$$

Eqs. (10) to (13) are solved by posing them as a constrained multivariable function $Y(X)$ and solving it iteratively to find the set of variables $X^*$ that minimizes $Y$. This minimization problem is generally expressed as follows.

$$X^* = arg \min_{F_{min} \leq F(X) \leq F_{max}} Y(X) \tag{17}$$

where $X$ is a vector of variables that is updated in each iteration to minimize $Y$, $F(X)$ is a vector of functions describing the range in which $X^*$ can be found, $F_{min}$ and $F_{max}$ are constraints describing the lower bound and upper bound of $F(X)$ respectively. For a given $\dot{m}$, $Q_h$ and $A_v$, the steady flow characteristics in a heated channel is obtained by solving the minimization problem with the following parameters and constraints.

$$X = [P_i, \dot{Q}_i] \tag{18}$$

$$Y = \left(\left|\frac{P_i - P_e - \Delta P}{\Delta P}\right| + \left|\frac{\dot{Q}_h - \dot{Q}_i - \dot{Q}_{loss}}{\dot{Q}_{loss}}\right|\right) \tag{19}$$

$$P_i \geq P_e, \quad \dot{Q}_i \leq \dot{Q}_h, \quad Y \leq 10^{-4} \tag{20}$$

The iterative procedure shown in Figure 3, begins with a first guess of the input variables $P_i$ and $\dot{Q}_i$. These values are used to compute $h_e$, $l_{ph}$, $T_{ph}$, $H_{ph}$ and $\Delta P$, which are then used to compute $T_w$, $\dot{Q}_{loss}$, and consequently $Y$. This procedure is repeated until an iteration step is smaller than the step size tolerance of the optimization tool and $Y \leq 10^{-4}$, while the input variables are updated with each iteration.

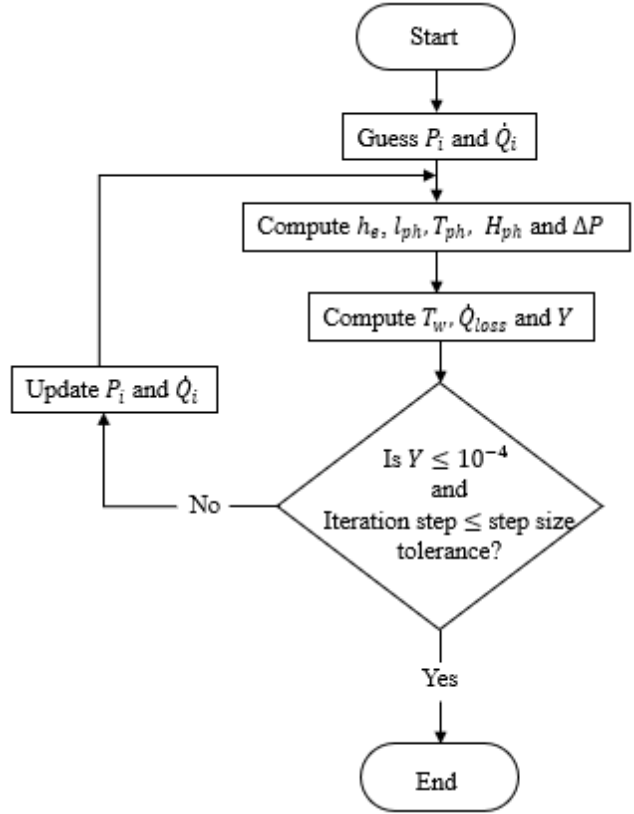

Figure 3. Iteration process for steady solution in a single channel.

In the case of a two parallel channel system with given heat loads $Q_{h1}$ and $Q_{h2}$, valve openings $A_{v_1}$ and $A_{v_2}$, and flow rate $\dot{m}$, the steady flow characteristics are obtained by solving the following minimization problem.

$$X = [P_i, \dot{Q}_{i,1}, \dot{Q}_{i,2}, \dot{m}_1^*] \tag{21}$$

$$\dot{m}_2^* = 1 - \dot{m}_1^* \tag{22}$$

$$Y = \frac{1}{min(\Delta P_1, \Delta P_2)}\sum_{j=1}^{2}|P_i - P_e - \Delta P_j| + \frac{1}{min(\dot{Q}_{i,1}, \dot{Q}_{i,2})}\sum_{j=1}^{2}|\dot{Q}_{h,j} - \dot{Q}_{i,j} - \dot{Q}_{loss,j}| \tag{23}$$

$$P_i \geq P_e,\ \dot{Q}_{i,1} \leq \dot{Q}_{h,1},\ \dot{Q}_{i,2} \leq \dot{Q}_{h,2},\ \dot{m}_1^* \leq 1,\ \lambda \leq \lambda_{max},\ Y \leq 10^{-4} \tag{24}$$

Here $\dot{m}_1^*$ and $\dot{m}_2^*$ are the flow fractions in channels 1 and 2, respectively. $\lambda$ is a linear stability criterion obtained from the maximum real part of the eigenvalues of the Jacobian matrix of $\frac{d}{dt}\begin{bmatrix}\dot{m}_1^*\\\dot{m}_2^*\end{bmatrix}$ [19], and $\lambda_{max}$ is the upper bound on $\lambda$. If $\lambda < 0$, the static solution is stable, and if $\lambda > 0$ the solution is unstable. Therefore, to obtain only stable solutions $\lambda_{max} = 0$. The iterative procedure for a two-parallel-channel system is similar to the single channel. However, there is an additional stability constraint ($\lambda$), as shown in Figure 4.

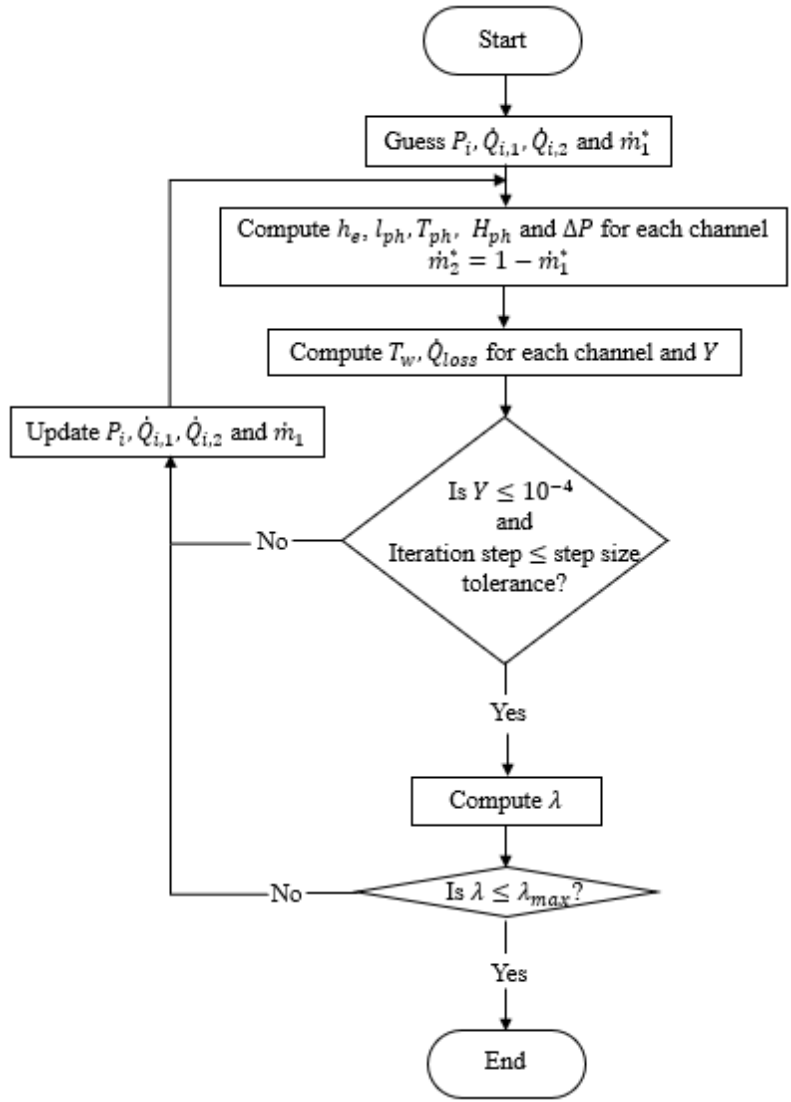

Figure 4. Iteration process for steady solution in two parallel channels.

### 2.2.2 Transient Model

The unsteady momentum balance equation (Eq. (2)) applied to a two-parallel-channel system allows for predicting the evolution of flow fractions, $\dot{m}_1^*$ and $\dot{m}_2^*$.

$$\frac{d}{dt}\begin{bmatrix} \dot{m}_1^* \\ \dot{m}_2^* \end{bmatrix} = \frac{L}{\dot{m}} \begin{bmatrix} \frac{1}{A_{cs,1}}(P_i - P_e - \Delta P_1) \\ \frac{1}{A_{cs,2}}(P_i - P_e - \Delta P_2) \end{bmatrix} \qquad (25)$$

Here $A_{cs,1}$ and $A_{cs,2}$ are the cross-sectional areas of channels 1 and 2 respectively. $P_i$ is obtained from the unsteady momentum balance equation for the whole assembly.

$$\frac{\dot{m}_{t+\Delta t} - \dot{m}_t}{\Delta t} = \sum_{j=1}^{2} \frac{L}{A_{cs,j}}(P_i - P_e - \Delta P_j) \qquad (26)$$

Here $\Delta t$ is the time step applied for estimating $d\dot{m}/dt$ numerically. The unsteady momentum balance equation has the most significant influence on the transient evolution of flow distribution in a two-parallel-channel assembly relative to the unsteady mass and energy conservation equations. Hence, in this study, the transient model consists of solving the Eqs. (10), (12), (13) and (25).

## 2.3 Flow Distribution and Entropy Generation

The static solution for different flow rates in a heated channel produces characteristic 'N' curves (red and black lines), as shown in Figure 5. For a given $\dot{m}$ in two parallel channels, the steady flow distribution solution $\dot{m}_1$ and $\dot{m}_2$ each lie on the characteristic curve corresponding to channel 1 and channel 2, respectively. From Figure 5, a fixed $\dot{m}$ may yield multiple flow distributions (1, 2 and 3). Linear stability analysis of these solutions[6], [8] would indicate that 1 and 3 are stable and feasible, while 2 is unstable, leaving us with two stable maldistributed flow solutions. We conduct an entropy analysis of each solution to identify the most feasible solution from these "stable" flow distributions.

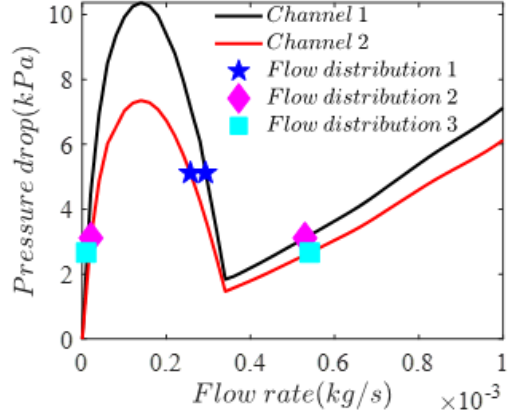

Figure 5. Steady flow distribution solutions in a two-parallel-channel assembly. Lines represent single-channel characteristic pressure curves for channels 1 (diameter $1.4 \times 10^{-3}$ m and heat load 60 W) and 2 (diameter $1.5 \times 10^{-3}$ m and heat load 60 W), respectively. The markers represent different mathematical solutions for flow distribution with a total flow rate of $5.5 \times 10^{-4}$ kg/s.

Entropy analysis is an effective tool for determining the direction of physical processes. For a process to be feasible, the rate of entropy generated ($\dot{S}_{gen}$) during that process must be greater than 0. In a system of $N$ parallel channels, the rate of entropy generation $\dot{S}_{gen}$ at steady state is given by the following equation.

$$\dot{S}_{gen} = \dot{m}\big(s_e(P_e, h_e) - s_i(P_i, T_i)\big) - \sum_{j=1}^{N} \frac{\dot{Q}_{i,j}}{T_{w,j}} \tag{27}$$

The specific entropies at the inlet ($s_i$) and outlet ($s_e$) of the channel assembly are functions of inlet pressure $P_i$ and temperature $T_i$, and exit pressure $P_e$ and specific enthalpy $h_e$, respectively. Entropy generated within a heated parallel channel assembly consists of entropy generated within each channel flow stream, entropy generated from splitting the flow at the common inlet, and entropy generated from mixing the flow at the common exit of the network. $\dot{S}_{gen}$ can be expressed using the following equation.

$$\dot{S}_{gen} = \dot{S}_{gen,mix} + \sum_{j=1}^{N} \dot{S}_{gen,j} \tag{28}$$

Here $\dot{S}_{gen,j}$ is the entropy generation rate within each channel of the parallel network. $\dot{S}_{gen,mix}$ is the rate of entropy generated by heat transfer and expansion corresponding to fluid emerging from each channel and mixing at the common exit. For an adiabatic mixing process at the common header, $\dot{S}_{gen,mix}$ is a function of the flow rate distribution $\dot{m}_j^* = \frac{\dot{m}_j}{\dot{m}}$ and the heat load distribution $\dot{Q}_{h,j}^* = \frac{\dot{Q}_{h,j}}{\dot{Q}_h}$.

$$\dot{S}_{gen,mix} = \dot{m}\left(s_{mix} - \sum_{j=1}^{N} \dot{m}_j^* s_{e,j}(P_e, h_{e,j})\right) \quad (29)$$

$$h_{e,j} = h_i + \frac{\dot{Q}_{h,j}^* \dot{Q}_{h,j}}{\dot{m}_j^* \dot{m}} \quad (30)$$

## 3   EXPERIMENTAL TESTBED

An experimental testbed consisting of a heated tank, gear pump, electronic valve, and evaporator assembly was constructed (Figure 6) to validate the static and dynamic models. The evaporator assembly consists of two capillary steel tubes with an internal diameter of $1.4 \times 10^{-3}$ m, outer diameter $3.175 \times 10^{-3}$ m and length $3.048 \times 10^{-1}$ m, wrapped with 125 W rope heaters and an outer layer of fiberglass insulation. Coupled to the ends of each steel tube are stop valves and flow meters (Omega FLR-1008ST, $\pm 3.33 \times 10^{-5}$ kg/s). Four thermocouples (Omega T-type, $\pm 1°C$) are attached to the wall of each tube at equidistant locations to monitor the wall temperature. Pressure sensors (Omega PX309-030A5V, $\pm 0.52$ kPa) and additional thermocouples are positioned at the inlet and exit of the assembly to monitor flow properties at these locations.

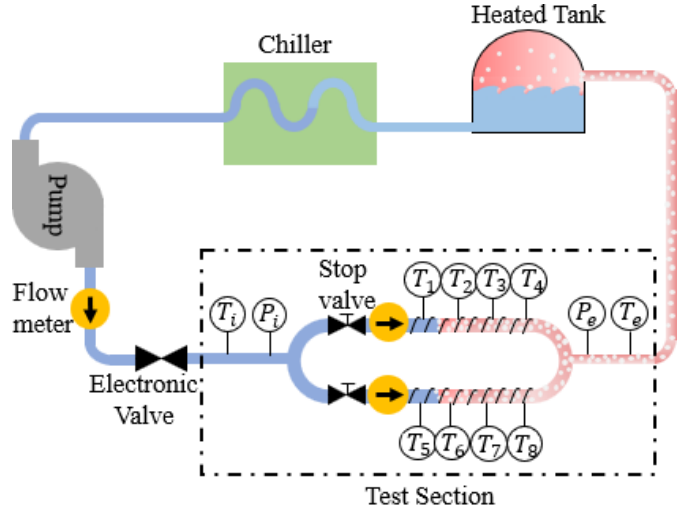

Figure 6. The experimental testbed to study static and dynamic characteristics of flow maldistbution.

## 4 RESULTS

### 4.1 Model Validation

In Figure 7a, a heat load of 70 W is supplied to each channel while fluid is supplied through the inlet header to the parallel channel assembly. Initially, at a high flow rate, the flow is uniformly distributed among each channel (Figure 7a-i) while $\dot{S}_{gen}$ decreases with $\dot{m}$ (Figure 7a-ii). However, as $\dot{m}$ is gradually decreased, the flow becomes severely maldistributed and $\dot{S}_{gen}$ suddenly increases. After flow maldistribution, $\dot{S}_{gen}$ decreases with subsequent decrease in $\dot{m}$.

For the transient experiment, channel 1 and channel 2 are supplied with a steady heat load of 40 W and 60 W, respectively, while the assembly is initially supplied with a flow rate $\dot{m}$ of about $10^{-3}$ kg/s as shown in Figure 7b-i. Like the static experiment, initially, the flow is equally distributed in the two channels. However, when the $\dot{m}$ is abruptly decreased to $4.6 \times 10^{-3}$ kg/s, the flow becomes severely maldistributed, with channel 1, in this case, receiving the bulk of the flow and channel 2 experiencing almost no flow, as shown in Figure 7b-ii. Our static and transient model predictions generally agree with experimental data[20].It should be noted that $\dot{Q}_{loss}$ is neglected in further application of this model.

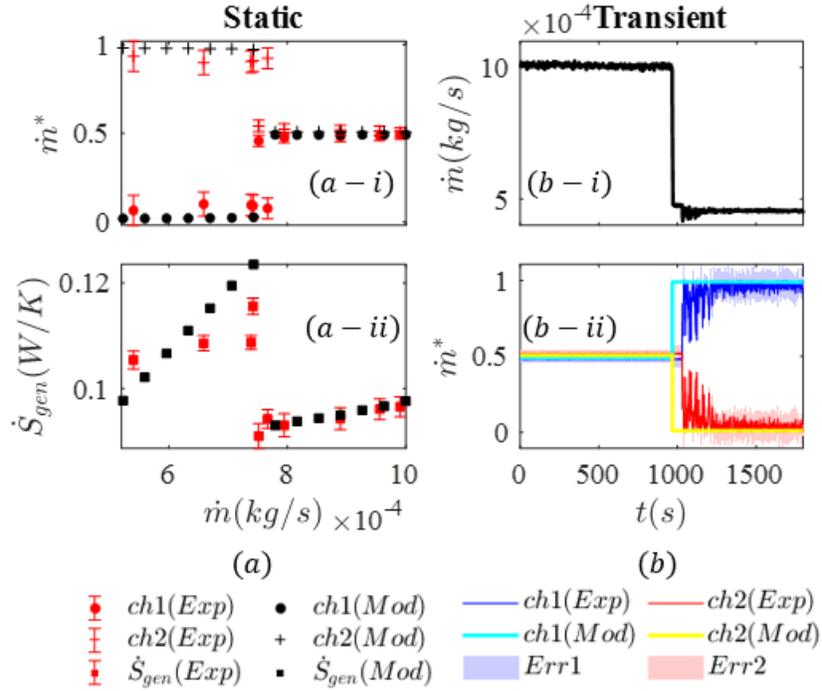

Figure 7. Model validation for both static and dynamic conditions

## 4.2 Entropy generation in a single channel

Flow boiling in a single channel is characterized by thermal and hydraulic resistances. These resistances are sources of irreversibility and contribute significantly to entropy production within the channel. Hydraulic resistance, primarily due to frictional forces and fluid expansion, causes a pressure drop $\Delta P = P_i - P_e$ as flows takes place through the channel. Likewise, thermal resistance requires a temperature difference, $\Delta T_w = T_w - T_f$ for heat transfer between the wall and the fluid. Hence, $\Delta P$ and $\Delta T_w$ denote a departure from ideal flow and heat transfer with no entropy generation and is a measure of irreversibility from the heat transfer and flow process, respectively.

Figure 8 shows the variation of $\Delta P$, $\Delta T_w$, and $\dot{S}_{gen}$ with $\dot{m}$, for different combinations of heat loads $\dot{Q}_h$ and channel internal diameter $D$ for a single channel.

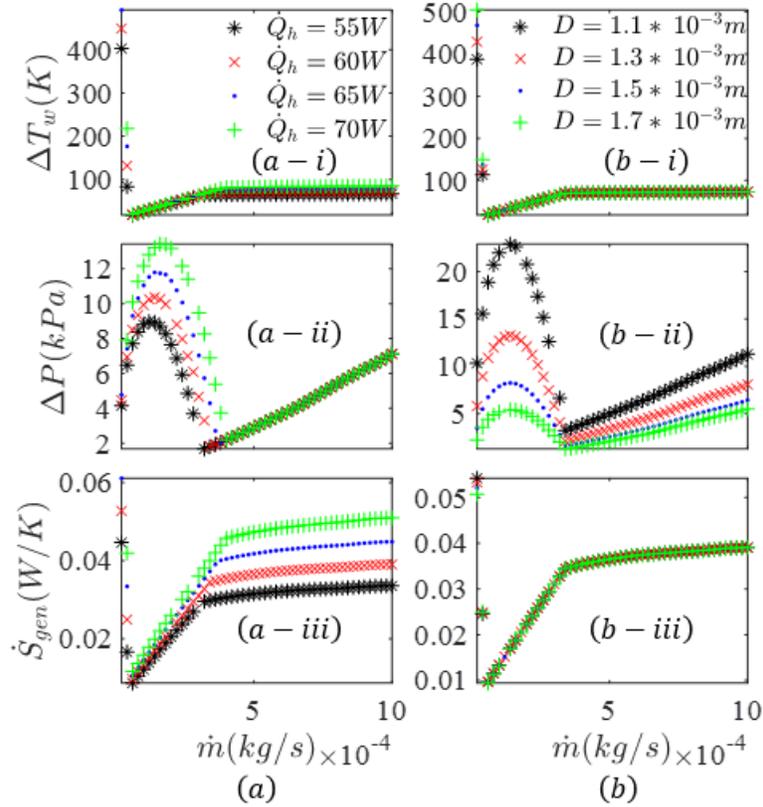

Figure 8. Variation of single-channel steady-state flow properties with (a) varying $\dot{m}$ and $\dot{Q}_h$ at $D = 1.4 \times 10^{-3}$ m (b) varying $\dot{m}$ and $D$ at $\dot{Q}_h = 60$ W

A large $\dot{m}$ generally corresponds to single-phase liquid at the channel exit, indicated by a positive slope of curves on the far right of each plot. As $\dot{m}$ decreases, the fluid at the channel exit becomes a two-phase vapor-liquid mixture, resulting in a change in slope from positive to negative in the $\Delta P$ versus $\dot{m}$ curves. The slope of the curves for other parameters remains positive but becomes steeper. Very small $\dot{m}$ on the far left of each plot corresponds to superheated vapor flow at the channel exit. In this region, the slope of the $\Delta P$ versus $\dot{m}$ curves become positive again, while the slope of the other curves changes from positive to a very steep negative value due to the occurrence of critical heat flux (CHF).

The variation in $\dot{Q}_h$ has no significant impact on $\Delta P$ for single-phase liquid at the channel exit (Figure 8a-ii). However, $\Delta P$ increases with an increase in $\dot{Q}_h$ when two-phase mixtures and superheated vapor exit the channels. The variation in $D$ results in significant variation in $\Delta P$, with $\Delta P$ increasing for smaller $D$ (Figure 8a-ii). An increase in $\dot{Q}_h$ increases $\Delta T_w$ and $\dot{S}_{gen}$ (Figure 8a-i and Figure 8a-iii). while variations in $D$ has no significant impact on $\Delta T_w$ and $\dot{S}_{gen}$ (Figure 8b-i and Figure 8b-iii). Generally, the variation in $\Delta T_w$ and $\dot{S}_{gen}$ versus $\dot{m}$ are similar and differ from the trends in $\Delta P$ versus $\dot{m}$, indicating that the dominant contribution towards irreversibility is associated more with heat transfer than flow characteristics.

## 4.3 Entropy generation due to adiabatic mixing of fluid

The fraction of flow and heat load in each channel influences $\dot{S}_{gen,mix}$. Figure 9 describes the variation of $\dot{S}_{gen,mix}$ with the variation in the fraction $\dot{m}_j^* = \frac{\dot{m}_j}{\dot{m}}$ of the total flow rate $\dot{m}$ in channel $j$, and fraction $\dot{Q}_{h,j}^* = \frac{\dot{Q}_{h,j}}{\dot{Q}_h}$ of the total heating $\dot{Q}_h$ on channel $j$ of a two parallel channel assembly. Therefore, for uniformly distributed flow and heating power, $\dot{m}_j^* = \dot{Q}_{h,j}^* = 0.5$.

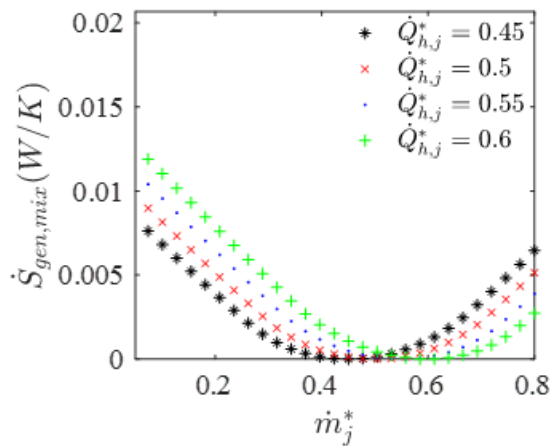

Figure 9. Variation of $\dot{S}_{gen,mix}$ with $\dot{m}_j^*$ and $\dot{Q}_{h,j}^*$ for $\dot{m} = 10^{-3}$kg and $\dot{Q}_h = 100$W

The $\dot{S}_{gen,mix}$ corresponding to a given $\dot{Q}_{h,j}^*$ is typically a parabola with the minimum $\dot{S}_{gen,mix}$ at $\dot{m}_j^* = \dot{Q}_{h,j}^*$. At this point, the thermal energy content of both fluid streams is the same, and hence, there is no entropy generation due to the irreversible fluid-to-fluid heat transfer during the mixing process. At points where $\dot{m}_j^* < \dot{Q}_{h,j}^*$, $\dot{S}_{gen,mix}$ increases with an increase in $\dot{Q}_{h,j}^*$, and at points where $\dot{m}_j^* > \dot{Q}_{h,j}^*$, $\dot{S}_{gen,mix}$ increases with a decrease in $\dot{Q}_{h,j}^*$. This variation shows that the maximum $\dot{S}_{gen,mix}$ occurs when the flow is highly maldistributed (far right or far left regions of the plot), corresponding to a high heat load with a low flow rate or low heat load with a high flow rate.

## 4.4 Entropy generation in a two-parallel channel assembly

Figure 10 describes the variation of $\Delta P, \Delta T_{w,1}, \Delta T_{w,2}, \dot{m}_1^*, \dot{m}_2^*$, and $\dot{S}_{gen}$ with $\dot{m}$ in the two-channel assembly. Initially, $\dot{m}$ is almost uniformly distributed among the channels, $\Delta T_{w,1}$ and $\Delta T_{w,2}$ are constant while $\Delta P$ and $\dot{S}_{gen}$ decrease as $\dot{m}$ decreases. With a further decrease in $\dot{m}$, phase change in the working fluid triggers severe flow maldistribution, characterized by increased flow in channel 1 and decreased flow in channel 2. This results in a jump in $\Delta T_{w,2}$ and $\Delta P$ across the assembly, with no change in $\Delta T_{w,1}$. The onset of CHF in channel 2 and increased hydraulic resistance in channel 1, causes $\dot{S}_{gen}$ to suddenly increase with flow maldistribution. After the onset of flow maldistribution, the influence of increasing $\Delta T_{w,2}$ causes $\dot{S}_{gen}$ to increase until the influence of decreasing $\dot{m}$ becomes dominant again causing $\dot{S}_{gen}$ to decrease.

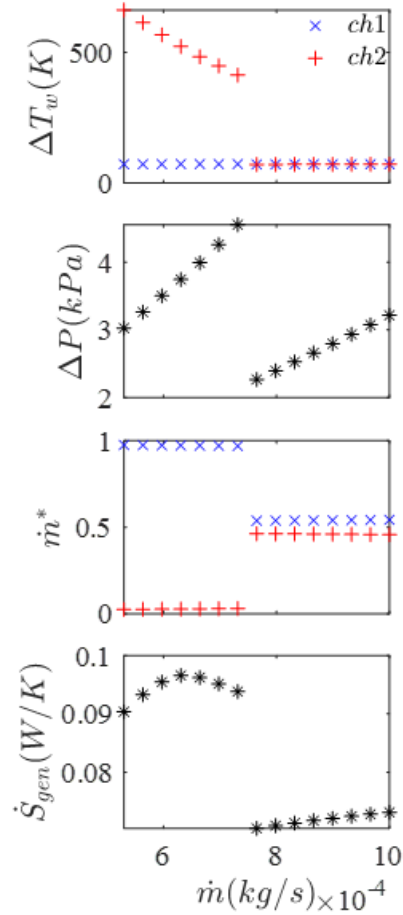

Figure 10. Effect of flow maldistribution on flow properties of a two parallel channel assembly with channel 1 characteristics corresponding to $D_1 = 1.4 \times 10^{-3}$ m , $\dot{Q}_{h,1} = 60$ W, and $A_{v_1} = 100$ % and channel 2 characteristics corresponding to $D_2 = 1.4 \times 10^{-3}$ m , $\dot{Q}_{h,2} = 60$ W, and $A_{v_2} = 50$ %.

## 4.5  Entropy generation and stability of flow distribution

Typically, multiphase flow distribution solutions in parallel channels are not unique (Figure 5). Modeling could indicate multiple solutions with both stable and unstable flow distribution for the same operating condition. As described earlier, a stability criterion $\lambda$ derived from the linear perturbation theory, is used to distinguish between stable ($\lambda < 0$) and unstable ($\lambda > 0$) flow distributions as shown in Figure 11a-i and Figure 11b-i . To provide a thermodynamic perspective to stability of flow distributions, Figure 11 compare the $\dot{S}_{gen}$ for stable flow distribution (black) with an unstable flow distribution profile (green).

In Figure 11a, the channel properties are identical except for the heat loads with $\dot{Q}_{h,1} = 50W$ and $\dot{Q}_{h,2} = 70W$. For large $\dot{m}$, flow solutions are uniformly distributed and stable while for small $\dot{m}$, the predicted flow distributions can be either severely maldistributed and stable or moderately non-uniform and unstable as shown in Figure 11a-i and Figure 11a-ii. In Figure 11b, the channel properties are identical except for the valve openings that are set at $A_{v_1} = 100\%$ and $A_{v_2} = 50\%$. For large $\dot{m}$, the flow solutions are stable and slightly unequally distributed while for small $\dot{m}$ the flow distribution is either severely maldistributed and stable or is uniform and unstable as shown in Figure 11b-i and Figure 11b-ii. In both Figure 11a-iii and Figure 11b-iii, the $\dot{S}_{gen}$ associated with the stable maldistributed flow solutions is greater than the $\dot{S}_{gen}$ associated with the moderately non-uniform or the uniformly distributed unstable flow solutions. Therefore, a severely maldistributed flow is thermodynamically preferred over other flow distributions satisfying the system constraints. Such severe maldistributions have been observed to occur, as shown in Figure 7.

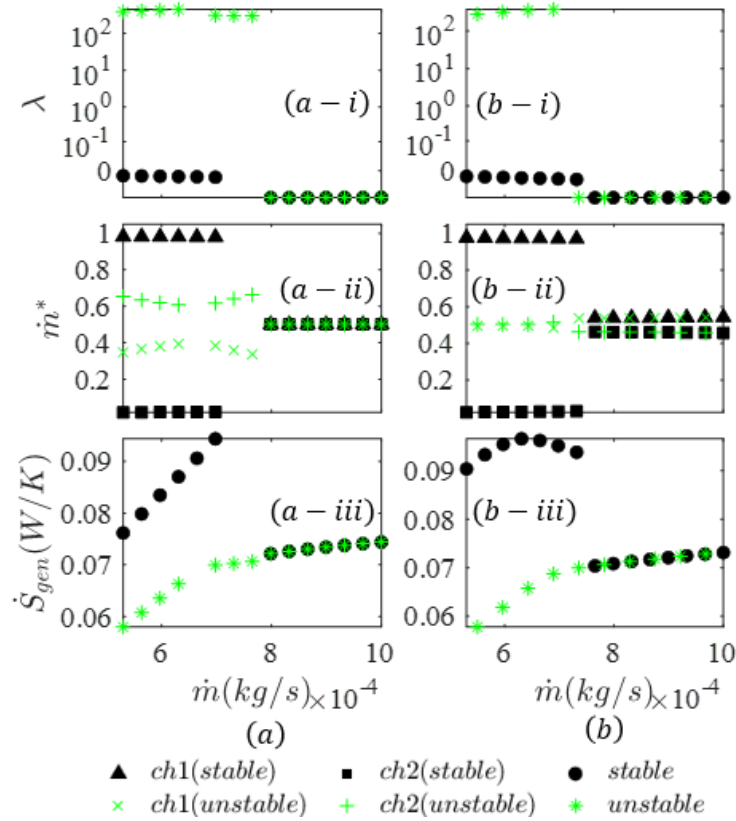

Figure 11. Comparison between stable and unstable flow distribution profile for : (a) $D_1 = D_2 = 1.4 \times 10^{-3} m$, $Av_1 = Av_2 = 100\%$, $\dot{Q}_{h,1} = 50W$ and $\dot{Q}_{h,2} = 70W$. (b) $D_1 = D_2 = 1.4 \times 10^{-3} m$, $\dot{Q}_{h,1} = \dot{Q}_{h,2} = 60W$, $Av_1 = 100\%$ and $Av_2 = 50\%$

## 4.6 Entropy generation and maldistributed flow solutions

Based on linear stability characteristics described in Figure 11, maldistributed flow solutions are inherently stable but could be non-unique as shown in Figure 5. This limits the application of the linear stability analysis since it only distinguishes between stable and unstable flow distribution solutions. For a given condition in a two-parallel channel system with perfectly identical individual channels, maldistributed flow solutions are mirror images of each other, with the flow magnitudes reversed compared with each other. These solutions are also indistinguishable when considering extensive thermodynamic properties (like $\dot{S}_{gen}$) associated with each solution. With the introduction of non-uniformities to the properties of the individual channels, as expected in practical applications, entropy analysis can determine the feasibility of each solution as the final state of a simple flow maldistribution process.

Let us consider a flow maldistribution process in a perfectly insulated $(\dot{Q}_i = \dot{Q}_h)$ two-parallel-channel system, with the following steady properties: flowrate $(\dot{m})$, identical heat loads $(\dot{Q}_{h,1} = \dot{Q}_{h,2} = \dot{Q}_h/2)$, and identical internal diameters $(D_1 = D_2)$. If the entropy generation rate of the final maldistributed state is $\dot{S}_{gen,0}$, since the individual channels are identical, flow maldistributions $\dot{m}_1 < \dot{m}_2$ and $\dot{m}_1 > \dot{m}_2$ are equivalent with respect to $\dot{S}_{gen,0}$. From Eq. 28,

$$\dot{S}_{gen,0} = \dot{S}_{gen,1} + \dot{S}_{gen,2} + \dot{S}_{gen,mix} \tag{31}$$

where $\dot{S}_{gen,1}$ and $\dot{S}_{gen,2}$ are the rate of entropy generation in channels 1 and 2, respectively. Introducing non-uniformities in channel heating rates and dimensions while maintaining a constant flow rate and total heat load will result in a change in the entropy generation rate by $d\dot{S}_{gen,0}$. The final rate of entropy generation $\dot{S}_{gen}$ is given by.

$$\dot{S}_{gen} = \dot{S}_{gen,0} + d\dot{S}_{gen,0} \tag{32}$$

$d\dot{S}_{gen,0}$ can be expressed using the chain rule.

$$d\dot{S}_{gen,0} = \sum_{j=1}^{2} \left( \frac{\partial \dot{S}_{gen,j}}{\partial \dot{m}_j} d\dot{m}_j + \frac{\partial \dot{S}_{gen,j}}{\partial \dot{Q}_{h,j}} d\dot{Q}_{h,j} + \frac{\partial \dot{S}_{gen,j}}{\partial D_j} dD_j + \frac{\partial \dot{S}_{gen,mix}}{\partial \dot{Q}_{h,j}^*} d\dot{Q}_{h,j}^* + \frac{\partial \dot{S}_{gen,mix}}{\partial \dot{m}_j^*} d\dot{m}_j^* \right) \tag{33}$$

The conservation of mass and energy equations become.

$$d\dot{m}_1 = -d\dot{m}_2 \tag{34}$$

$$d\dot{Q}_{h,1} = -d\dot{Q}_{h,2} \tag{35}$$

From the momentum balance equation, $\dot{m}_j$ is a function of $\dot{Q}_{h,j}$, $D_j$ and the common pressure drop $\Delta P$ as shown in Figure 8a-i and Figure 8b-i. $d\dot{m}_j$ can be expressed using the chain rule.

$$d\dot{m}_j = \frac{\partial \dot{m}_j}{\partial \Delta P} d\Delta P + \frac{\partial \dot{m}_j}{\partial \dot{Q}_{h,j}} d\dot{Q}_{h,j} + \frac{\partial \dot{m}_j}{\partial D_j} dD_j \tag{36}$$

Based on the trends in Figure 8 and Figure 9 the characteristics of the partial derivative terms in Eq. 33 and Eq. 36 for a maldistributed flow in a two parallel channel is summarized in Table 1.

**Table 1**
Effect of system parameters on $\dot{S}_{gen,j}$ and $\dot{m}_j$

| | Reference Figures | $\dot{m}_j \to \dot{m}$ | $\dot{m}_j \to 0$ |
|---|---|---|---|
| $\frac{\partial \dot{S}_{gen,j}}{\partial \dot{m}_j}$ | 8a-iii or 8b-iii | $> 0$ | $\ll 0$ |
| $\frac{\partial \dot{S}_{gen,j}}{\partial \dot{Q}_{h,j}}$ | 8a-iii | $> 0$ | $> 0$ |
| $\frac{\partial \dot{S}_{gen,j}}{\partial D_j}$ | 8b-iii | $\approx 0$ | $\approx 0$ |
| $\frac{\partial \dot{S}_{gen,mix}}{\partial \dot{Q}_{h,j}^*}$ | 9 | $< 0$ | $> 0$ |
| $\frac{\partial \dot{S}_{gen,mix}}{\partial \dot{m}_j^*}$ | 9 | $> 0$ | $< 0$ |
| $\frac{\partial \dot{m}_j}{\partial P_d}$ | 8a-i or 8b-i | $> 0$ | $> 0$ |
| $\frac{\partial \dot{m}_j}{\partial \dot{Q}_{h,j}}$ | 8a-i | $\approx 0$ | $< 0$ |
| $\frac{\partial \dot{m}_j}{\partial D_j}$ | 8b-i | $> 0$ | $> 0$ |

### 4.6.1 Variation in diameter

Let non-uniformity be introduced by slightly changing the diameter of one channel, Eq. 33 reduces to

$$d\dot{S}_{gen,0} = \sum_{j=1}^{2} \left( \frac{\partial \dot{S}_{gen,j}}{\partial \dot{m}_j} d\dot{m}_j + \frac{\partial \dot{S}_{gen,j}}{\partial D_j} dD_j + \frac{\partial \dot{S}_{gen,mix}}{\partial \dot{m}_j^*} d\dot{m}_j^* \right) \tag{37}$$

The conservation of mass in Eq. 34 based on Eq. 36 becomes.

$$\frac{\partial \dot{m}_1}{\partial \Delta P} d\Delta P + \frac{\partial \dot{m}_1}{\partial D_1} dD_1 = -\left( \frac{\partial \dot{m}_2}{\partial \Delta P} d\Delta P + \frac{\partial \dot{m}_2}{\partial D_2} dD_2 \right) \tag{38}$$

If $D_1$ is increased by $dD$ and the maldistributed flow is such that $\dot{m}_1 \approx \dot{m}$ and $\dot{m}_2 \approx 0$, then $dD_1 > 0$, $dD_2 = 0$. From Table 1, Eq.38 is only satisfied if $d\Delta P < 0$ in this case, which implies $d\dot{m}_1 > 0$ and $d\dot{m}_1^* > 0$ while $d\dot{m}_2 < 0$ and $d\dot{m}_2^* < 0$. Based on the magnitude of the partial derivative terms in Table 1 with respect to channel 1 and channel 2, we can deduce that all the $\frac{\partial}{\partial}d$ terms in Eq. 37 are $\geq 0$ for this case, implying $d\dot{S}_{gen,0} > 0$.

If $D_2$ is decreased by $dD$ and the maldistributed flow is such that $\dot{m}_1 \approx 0$ and $\dot{m}_2 \approx \dot{m}$, then $dD_1 = 0$, $dD_2 < 0$. From Table 1, Eq.38 is only satisfied if $d\Delta P > 0$ in this case, which implies $d\dot{m}_1 > 0$ and $d\dot{m}_1^* > 0$ while $d\dot{m}_2 < 0$ and $d\dot{m}_2^* < 0$. Based on the magnitude of the partial derivative terms in Table 1 with respect to channel 1 and channel 2, we can deduce that all the $\frac{\partial}{\partial}d$ terms in Eq. 37 are $\leq 0$ for this case, implying $d\dot{S}_{gen,0} < 0$.

Therefore, for a case of a maldistributed flow in a two parallel channel system, where $\dot{Q}_{h,1} = \dot{Q}_{h,2}$ and $D_1 > D_2$, $\dot{S}_{gen}$ corresponding to $\dot{m}_1 > \dot{m}_2$ is greater than $\dot{S}_{gen}$ corresponding to $\dot{m}_1 < \dot{m}_2$. Based on the magnitude of $\dot{S}_{gen}$ for the given system conditions $\dot{m}_1 > \dot{m}_2$ is thermodynamically more favorable and is likely the final maldistributed state in the process of flow maldistribution. This implies that when flow maldistribution occurs in a two parallel channel assembly with varying internal diameters and identical heat loads, flow is concentrated in the channel with larger diameter while that with the smaller diameter is starved of fluid. This outcome is corroborated in previous studies[5], [21].

### 4.6.2 Variations in heat load

If non-uniformity is introduced by adding $d\dot{Q}_h$ to channel 1 and removing $d\dot{Q}_h$ from channel 2. Therefore $d\dot{Q}_{h,1} > 0$ and $d\dot{Q}_{h,1}^* > 0$ while $d\dot{Q}_{h,2} < 0$ and $d\dot{Q}_{h,2}^* < 0$. Eq. 33 reduces to

$$d\dot{S}_{gen,0} = \sum_{j=1}^{2}\left(\frac{\partial \dot{S}_{gen,j}}{\partial \dot{m}_j}d\dot{m}_j + \frac{\partial \dot{S}_{gen,j}}{\partial \dot{Q}_{h,j}}d\dot{Q}_{h,j} + \frac{\partial \dot{S}_{gen,mix}}{\partial \dot{Q}_{h,j}^*}d\dot{Q}_{h,j}^* + \frac{\partial \dot{S}_{gen,mix}}{\partial \dot{m}_j^*}d\dot{m}_j^*\right) \quad (39)$$

The conservation of mass in Eq. 34 based on Eq. 36 becomes.

$$\frac{\partial \dot{m}_1}{\partial \Delta P}d\Delta P + \frac{\partial \dot{m}_1}{\partial \dot{Q}_{h,1}}d\dot{Q}_{h,1} = -\left(\frac{\partial \dot{m}_2}{\partial \Delta P}d\Delta P + \frac{\partial \dot{m}_2}{\partial \dot{Q}_{h,2}}d\dot{Q}_{h,2}\right) \quad (40)$$

If the maldistributed flow is such that $\dot{m}_1 \approx 0$ and $\dot{m}_2 \approx \dot{m}$. From Table 1, Eq. 40 is only satisfied if $d\Delta P > 0$ in this case, which implies $d\dot{m}_1 < 0$ and $d\dot{m}_1^* < 0$ while $d\dot{m}_2 > 0$ and $d\dot{m}_2^* > 0$. Based on the magnitude of the partial derivative terms in Table 1 with respect to channel 1 and channel 2, we can deduce that all the $\frac{\partial}{\partial}d$ terms in Eq. 39 are $> 0$ except $\frac{\partial \dot{S}_{gen,2}}{\partial \dot{Q}_{h,2}}d\dot{Q}_{h,2}$. However, from Table 1 $\left|\frac{\partial \dot{S}_{gen,1}}{\partial \dot{m}_1}\right| \gg \left|\frac{\partial \dot{S}_{gen,2}}{\partial \dot{Q}_{h,2}}\right|$, implying $d\dot{S}_{gen,0} > 0$.

If the maldistributed flow is such that $\dot{m}_1 \approx \dot{m}$ and $\dot{m}_2 \approx 0$. From Table 1, Eq. 40 is only satisfied if $d\Delta P < 0$ in this case, which implies $d\dot{m}_1 > 0$ and $d\dot{m}_1^* > 0$ while $d\dot{m}_2 < 0$ and $d\dot{m}_2^* < 0$. Based on the magnitude of the partial derivative terms in Table 1 with respect to channel 1 and channel 2, we can deduce that all the $\frac{\partial}{\partial}d$ terms in Eq. 39 are $< 0$ except $\frac{\partial \dot{S}_{gen,1}}{\partial \dot{Q}_{h,1}}d\dot{Q}_{h,1}$. However, from Table 1 $\left|\frac{\partial \dot{S}_{gen,2}}{\partial \dot{m}_2}\right| \gg \left|\frac{\partial \dot{S}_{gen,1}}{\partial \dot{Q}_{h,1}}\right|$, implying $d\dot{S}_{gen,0} < 0$.

Therefore, for a case of a maldistributed flow in a two parallel channel system where $\dot{Q}_{h,1} > \dot{Q}_{h,2}$ and $D_1 = D_2$, $\dot{S}_{gen}$ corresponding to $\dot{m}_1 < \dot{m}_2$ is greater than $\dot{S}_{gen}$ corresponding to $\dot{m}_1 > \dot{m}_2$. Based on the magnitude of $\dot{S}_{gen}$ for the given system conditions $\dot{m}_1 > \dot{m}_2$ is thermodynamically more favorable and is likely the final maldistributed state in the process of flow maldistribution. This implies that when flow maldistribution occurs in a two parallel channel assembly with varying heat loads and identical geometry, flow is concentrated in the channel with smaller heat load while that with the larger heat load is starved of fluid. This outcome is corroborated in previous studies[5], [21].

Based on conclusions from both scenarios, Figure 12 compares $\dot{S}_{gen}$ for expected maldistributed flow states (black) with $\dot{S}_{gen}$ for unlikely maldistributed flow states (green). All conditions are identical for both channels except the channel heat loads in Figure 12a or channel internal diameter in Figure 12b. Before the occurrence of flow maldistribution, the flow solutions (Figure 12a-i and Figure 12b-i) and their corresponding $\dot{S}_{gen}$ (Figure 12a-ii and Figure 12b-ii) are identical. However, after the occurrence of flow maldistribution, the expected flow distribution is characterized by small flow fraction in channel 1 and large flow fraction in channel 2 while the unlikely flow distribution is characterized by large flow fraction in channel 1 and small flow fraction in channel 2. A comparison of $\dot{S}_{gen}$ associated with both flow solutions shows that the $\dot{S}_{gen}$ from the expected stable flow distribution is greater than $\dot{S}_{gen}$ from the unlikely but stable flow distribution.

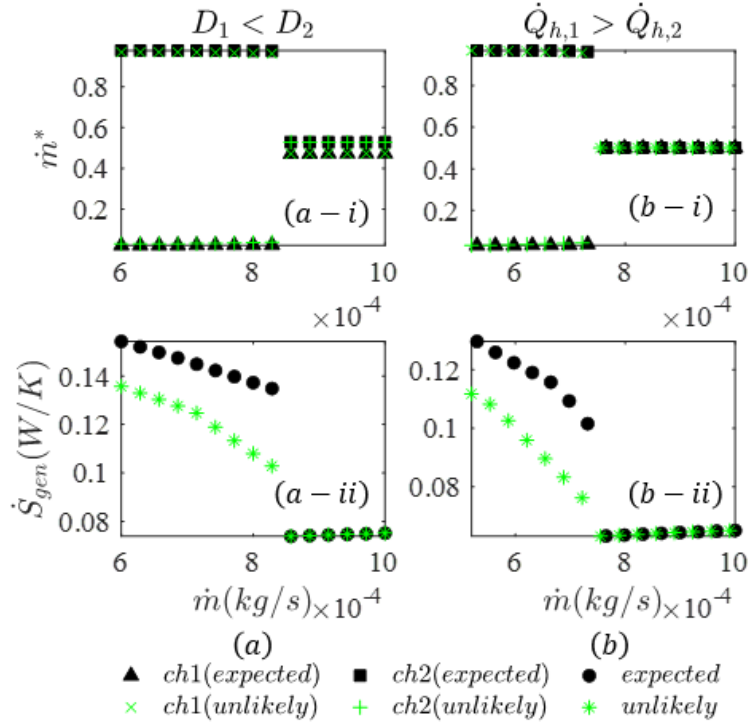

Figure 12. Comparison between expected and unlikely maldistributed flow states in a two parallel channel assembly for : (a) $D_1 = 1.3 \times 10^{-3}$ m , $D_2 = 1.3 \times 10^{-3}$ m and $\dot{Q}_{h,1} = \dot{Q}_{h,2} = 60$ W (b) $\dot{Q}_{h,1} = 60$ W, $\dot{Q}_{h,2} = 50$ W and $D_1 = D_2 = 1.4 \times 10^{-3}$ m

Figure 13 shows a parametric study to compare $\dot{S}_{gen}$ for maldistributed flow solutions with $\dot{m}_1 > \dot{m}_2$ (blue) and $\dot{m}_1 < \dot{m}_2$ (red) as a function of $D_1^* = \frac{D_1}{D_1+D_2}$ (left) and $\dot{Q}_{h,1}^* = \frac{Q_{h,1}}{Q_{h,1}+Q_{h,2}}$ (right). In Figure 13 (left), for solutions with $\dot{m}_1 > \dot{m}_2$, $\dot{S}_{gen}$ increases with increasing $D_1^*$. For solutions with $\dot{m}_1 < \dot{m}_2$, $\dot{S}_{gen}$ increases with decreasing $D_1^*$. This behavior is due to the increase in disparity between flow rates $|\dot{m}_1 - \dot{m}_2|$ and consequently CHF with increase in $D_1^*$ when $\dot{m}_1 > \dot{m}_2$ or decrease in $D_1^*$ when $\dot{m}_1 < \dot{m}_2$. At $D_1^* < 0.5$, solutions with $\dot{m}_1 < \dot{m}_2$ have a larger $\dot{S}_{gen}$ compared to solutions with $\dot{m}_1 > \dot{m}_2$. While, at $D_1^* > 0.5$, solutions with $\dot{m}_1 > \dot{m}_2$ have a larger $\dot{S}_{gen}$ compared to solutions with $\dot{m}_1 < \dot{m}_2$.

In Figure 13 (right), $\dot{S}_{gen}$ corresponding to both solutions with $\dot{m}_1 > \dot{m}_2$ and $\dot{m}_1 < \dot{m}_2$ increase with an increase in $\dot{Q}_{h1}^*$. This is caused by the increase in irreversibility associated with an increase in heat load. Also, as $\dot{Q}_{h1}^*$ increases the influence of $\dot{S}_{gen,mix}$ (Figure 9) becomes more dominant in solutions with $\dot{m}_1 < \dot{m}_2$ when compared to solutions with $\dot{m}_1 > \dot{m}_2$. As a result, at $\dot{Q}_{h1}^* < 0.5$, solutions with $\dot{m}_1 > \dot{m}_2$ have a larger $\dot{S}_{gen}$ compared to solutions with $\dot{m}_1 < \dot{m}_2$. While, at $\dot{Q}_{h1}^* > 0.5$, solutions with $\dot{m}_1 < \dot{m}_2$ have a larger $\dot{S}_{gen}$ compared to solutions with $\dot{m}_1 > \dot{m}_2$. Based on the solution corresponding to the maximum $\dot{S}_{gen}$ in Figure 13 (left), $\dot{m}_1 < \dot{m}_2$ if $D_1^* < 0.5$ and $\dot{Q}_{h1}^* = 0.5$, while $\dot{m}_1 > \dot{m}_2$ if $D_1^* > 0.5$ and $\dot{Q}_{h1}^* = 0.5$. In Figure 13 (right), $\dot{m}_1 > \dot{m}_2$ if $\dot{Q}_{h1}^* < 0.5$ and $D_1^* = 0.5$, while $\dot{m}_1 < \dot{m}_2$ if $\dot{Q}_{h1}^* > 0.5$ and $D_1^* = 0.5$.

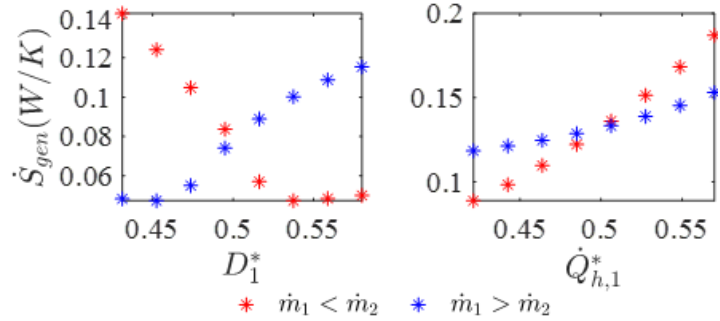

Figure 13. Comparison between $\dot{S}_{gen}$ corresponding to $\dot{m}_1 < \dot{m}_2$ and $\dot{m}_1 < \dot{m}_2$ maldistributed flow solutions for different values of $D_1^*$ and $\dot{Q}_{h1}^*$

## I. CONCLUSION

This study analyzes the relationship between two-phase flow distribution and entropy generation rate in a parallel channel assembly to address the challenge of a multiplicity of flow distribution solutions associated with the same conditions. The nonlinearity of the characteristic curves associated with two-phase flow in single channels indicates that stable theoretical solutions to flow distribution in a multi-channel network are often non-unique. In order to solve this challenge, previous studies applied linear stability analysis to determine the feasibility of a solution. However, this approach provides no underlying reason why a flow distribution is preferred over others, and it is limited in its applicability to distinguishing between stable and unstable flow distributions.

Therefore, we explore using an entropy analysis to predict the flow distribution in a two-parallel-channel network. In this study, entropy generation in parallel channel networks is divided into entropy generation within individual channels and entropy generation during the mixing of fluids at the common headers of the parallel channel network. The entropy analysis in a single channel with a constant heat load shows that hydraulic sources of irreversibility mainly drive entropy generation before the occurrence of CHF, while thermal sources of irreversibility become dominant after the occurrence of CHF. Also, entropy generation from mixing fluid at the common exit is a function of the disparity in the thermal content of each channel fluid stream. We show that entropy generation in a maldistributed flow is greater than any unstable flow distribution under the same conditions. Therefore, during phase change and within given system constraints, maldistributed flow is thermodynamically favored over other forms of flow distribution. However, this doesn't imply that system constraints may not be varied to ensure a more evenly distributed flow. Although maldistributed flow solutions are stable, these solutions are also non-unique. To distinguish between non-unique stable maldistributed flow solutions, we apply the trends observed from flow analyses in single channel and in common header of the parallel channel network to differential equations describing the change in entropy generation rate. Through this, we show that for a process of flow maldistribution under certain conditions, the final stable distribution with the highest rate of entropy generation is thermodynamically favored and will spontaneously occur. This is fundamental in understanding flow distribution in parallel channels and is applicable in optimizing the design of robust thermal systems against flow maldistribution.